\documentclass[twocolumn,showpacs,superscriptaddress,aps,prx,floatfix]{revtex4-2}
\usepackage{graphicx}
\usepackage{amsmath,amssymb}
\usepackage[utf8]{inputenc}
\usepackage{hyperref}
\usepackage{xcolor}
\hypersetup{colorlinks=true,citecolor={blue},linkcolor={blue},urlcolor={blue}}
\usepackage{units}
\usepackage[normalem]{ulem}
\usepackage{upgreek}
\usepackage{wasysym}
\usepackage{soul}

\makeindex

\begin{document}

\definecolor{lilac}{RGB}{150,100,180}
\definecolor{green}{RGB}{80,200,120}
\definecolor{orange}{RGB}{250,150,50}
\setstcolor{red}
\newcommand{\Stella}[1]{{\color{lilac}{#1}}}
\newcommand{\hs}[1]{\textcolor{red}{#1}}
\newcommand{\sa}[1]{{\color{green}{#1}}}
\newcommand{\jt}[1]{{\color{orange}{#1}}}

\newcommand\hcancel[2][red]{\setbox0=\hbox{$#2$}%
\rlap{\raisebox{.45\ht0}{\textcolor{#1}{\rule{\wd0}{1pt}}}}#2} 

\title{Solving the max-3-cut problem with coherent networks} 


\date{\today}

\author{S. L. Harrison}\email{S.L.Harrison@soton.ac.uk}
\affiliation{School of Physics and  Astronomy, University of Southampton, Southampton, SO17 1BJ, United Kingdom}

\author{H. Sigurdsson}
\affiliation{School of Physics and  Astronomy, University of Southampton, Southampton, SO17 1BJ, United Kingdom}
\affiliation{Skolkovo Institute of Science and Technology, Bolshoy Boulevard 30, bld. 1, Moscow, 121205, Russia}

\author{S. Alyatkin}
\affiliation{Skolkovo Institute of Science and Technology, Bolshoy Boulevard 30, bld. 1, Moscow, 121205, Russia}

\author{J. D. T\"{o}pfer}
\affiliation{School of Physics and  Astronomy, University of Southampton, Southampton, SO17 1BJ, United Kingdom}
\affiliation{Skolkovo Institute of Science and Technology, Bolshoy Boulevard 30, bld. 1, Moscow, 121205, Russia}

\author{P. G. Lagoudakis}\email{P.Lagoudakis@skoltech.ru}
\affiliation{Skolkovo Institute of Science and Technology, Bolshoy Boulevard 30, bld. 1, Moscow, 121205, Russia}
\affiliation{School of Physics and  Astronomy, University of Southampton, Southampton, SO17 1BJ, United Kingdom}

\begin{abstract}
	Many computational problems are intractable through classical computing and, as Moore's law is drawing to a halt, demand for finding alternative methods in tackling these problems is growing. Here, we realize a liquid light machine for the NP-hard max-3-cut problem based on a network of synchronized exciton-polariton condensates. We overcome the binary limitation of the decision variables in Ising machines using the continuous-phase degrees of freedom of a coherent network of polariton condensates. The condensate network dynamical transients provide optically-fast annealing of the XY Hamiltonian. We apply the Goemans and Williamson random hyperplane technique, discretizing the XY ground state spin configuration to serve as ternary decision variables for an approximate optimal solution to the max-3-cut problem. Applications of the presented coherent network are investigated in image-segmentation tasks and in circuit design.
\end{abstract}

\maketitle

\section{Introduction}

Complexity in nature is as widespread as it is diverse, and with the turn of the age there has been a rapid rise in non-Boolean strategies designed to tackle complex computational problems too cumbersome for conventional Turing-based computers which are limited by the {\it Von-Neumann Bottleneck}, as well as CMOS architectures approaching their limits. The need for new computational methods is underscored by many scientific fields such as those devoted to climate change~\cite{ibsen-jensen_computational_2015,zhang_energy_2016},
drug design~\cite{jayjaraj_drug_2016}, development of new materials and batteries
~\cite{pierce_protein_2002}, and so on, that are dependent on solving complex problems. A class of such problems is the computationally intractable NP-complete class, where an estimated solution can be verified in polynomial time, though no algorithm exists to calculate an exact solution. As these problems are so widely encountered, they are often approached using approximation algorithms or heuristic methods such as semi-definite programming~\cite{goemans_improved_1995,zhang_complex_2006}, genetic algorithms~\cite{such_deep_2018}, and nature inspired heuristic algorithms~\cite{yang_nature-inspired_2014,tovey_nature-inspired_2018}. However, instead of building an approximate optimization algorithm concerned with minimizing a cost function, an alternative is mapping the problem to a physical system that relaxes to the ground state of its energy landscape corresponding to the global minimum of the cost function. Coherent networks, which are being regarded as the next possible generation of both quantum~\cite{Aspuru-Guzik_NatPho2012} and classical computational devices~\cite{Sun_Nature2015}, have generated much interest with photonic-based classical annealers already realized for both Ising~\cite{marandi_network_2014, mcmahon_fully_2016, inagaki_coherent_2016, Inagaki_NatPho2016, Kyriienko_PRB2019, Pierangeli_PRL2019, Bohm_NatComm2019, luo_classical_2020, RoquesCarmes_NatCom2020} and XY spin Hamiltonians~\cite{nixon_observing_2013, berloff_realizing_2017,lagoudakis_polariton_2017, takeda_boltzmann_2017, gershenzon_exact_2020}, waveguide networks for the subset sum problem~\cite{Xu_ScieAdv2020}, and digital degenerate cavity laser for the phase retrieval problem~\cite{Tradonsky_SciAdv2019}. 

Here, we test the concept of a liquid light machine based on planar networks of exciton-polariton condensates in semiconductor microcavities and demonstrate their ability to optimize a maximum-3-cut computational problem compared to the brute force method with focus on two types of real world applications,  image-segmentation and constrained-via-minimization in circuit design.
Utilizing the recent developments connecting the dissipative nature of polariton condensate dynamics to minimization of the XY model~\cite{berloff_realizing_2017,Kalinin_SciRep2018, harrison_synchronization_2020}
we apply a random hyperplane technique to bin the XY ground state spins obtained from our polariton network into ternary decision variables~\cite{goemans_improved_1995, Goemans_JouCompSys2004}. We proceed to map the NP-hard max-3-cut (M3C) optimization problem~\cite{garey_computers_1979} to the energy minimization of a ternary phase-discretized XY model, which is then near-optimally solved using the decision variables obtained from the standard XY model. Variants on the max-cut problem have many applications, including social network modelling~\cite{harary_measurement_1959}, statistical physics~\cite{barahona_computational_1982}, portfolio risk analysis \cite{harary_signed_2002}, circuit design~\cite{barahona_application_1988}, image-segmentation~\cite{de_sousa_estimation_2013}, and more~\cite{barahona_network_1996, white_chapter_1973, deb_proceedings_2001}. 

Unlike Ising machines and quantum annealers, which can only be mapped to the max-2-cut problem~\cite{barahona_computational_1982,inagaki_coherent_2016,zhou_quantum_2020}, the continuous degree of freedom of the variables (spins) in XY systems makes their approximate partition into higher dimensional decision variables possible, with a direct mapping to the M3C problem. It also offers a perspective on whether higher order max-$k$-cuts can be approximated using such continuous phase systems. Our method can be applied to any system of interacting oscillators defined by their relative phases such as electronic circuits, laser systems, and condensates. However, polariton condensate systems have potential advantages over other light-based optimizers \cite{yamamoto_coherent_2017, hamerly_experimental_2019, gershenzon_exact_2020,vretenar_controllable_2021,reifenstein_coherent_2021,reddy_phase_2022} due to the continuous-phase-locking capabilities of the condensates, strong nonlinearities, the ability to arbitrarily control polariton graph geometries and coupling strengths  \cite{Alyatkin_PRL2020}, their long coherence length (reported up to 120~$\upmu$m \cite{topfer_engineering_2021}), their robust network synchronization resulting from the ballistic nature of the condensate interaction mechanism \cite{ohadi_nontrivial_2016,topfer_time-delay_2020}, plus the robustness of the inorganic semiconductor microcavity \cite{cilibrizzi_polariton_2014}.

%
\section{Mapping the M3C Problem to a Ternary XY Model} \label{sec.2}
The M3C problem is as follows: given an undirected graph $\mathcal{G} = (\mathcal{V},\mathcal{E})$ consisting of vertices $\mathcal{V}$ and weighted edges $\mathcal{E}$, the M3C is the partition of $\mathcal{V}$ into three subsets, such that the sum of all edge weights that connect between different subsets (labelled the ``weight of cut'') is maximized. Following similar arguments presented by Barahona for an Ising spin system~\cite{barahona_application_1988} it can be shown that a map from maximization of a M3C problem to minimizing the energy of a ternary XY spin system,
\begin{equation}
H_T = -\sum_{ij}J_{ij}\mathbf{s}_i \cdot \mathbf{s}_j,
\label{eqn:hamiltonain1}
\end{equation}
where
\begin{align} \label{m3c.spins}
\mathbf{s} &=   \begin{pmatrix}
\cos(\theta) \\
\sin(\theta)\\
\end{pmatrix}, \qquad  
\theta\in
\left\{
0,\frac{2\pi}{3},\frac{4\pi}{3}\right\}.
\end{align}
exists. 

Let us define an undirected graph $\mathcal{G} = (\mathcal{V},\mathcal{E})$ with edges $\mathcal{E}$ connecting vertices $\mathcal{V}$, where $\mathcal{V}$ contain the ternary spins $\mathbf{s}$ from Eq.~\eqref{m3c.spins}. Each edge connecting vertex $\mathcal{V}_i$ and $\mathcal{V}_j$, with corresponding spins $\mathbf{s}_i$ and $\mathbf{s}_j$, is assigned a weight $J_{ij} = J_{ji}$. We define three sets of spins corresponding to the three orientations of $\theta_i$ given by Eq.~\eqref{m3c.spins},
	\begin{equation}
	\mathcal{V}_n = \left\{ i\in \mathcal{V} \ | \ \theta_i = \frac{2\pi n}{3} \right\}, \qquad n=0,1,2.
	\end{equation}
	Let us define $\mathcal{E}_{n}$ as the set of edges connecting spins within each set $\mathcal{V}_{n}$ and $\delta \mathcal{E}$ as the set of edges connecting spins between different sets of vertices. We can then rewrite Eq.~\eqref{eqn:hamiltonain1} in the following manner,
	\begin{equation}
	H_T = -\sum_{i,j\in \mathcal{E}_{n}} J_{ij}+\frac{1}{2}\sum_{i,j \in \delta \mathcal{E}}J_{ij}.
	\end{equation}
	By defining the sum total of all the edges $C = \sum_{ij\in \mathcal{E}} J_{ij}$, we then have,
	\begin{equation}
	H_T + C = \frac32\sum_{i,j\in \delta \mathcal{E}}J_{ij}.
	\label{eq:weightOfCut}
	\end{equation}
	Since $C$ is invariant on the spin configuration, it can be seen that minimizing Eq.~\eqref{eqn:hamiltonain1} is the same as maximizing a M3C problem by redefining $J_{ij} = -\mathcal{W}_{ij}$, where $\mathcal{W}_{ij}$ is the weight of the edge connecting $\mathcal{V}_i$ and $\mathcal{V}_j$,	
	\begin{equation}
	\text{min}{\left[H_T \right]}\quad \leftrightarrow \quad \text{max}{\left[ \sum_{i,j \in \delta \mathcal{E}}\mathcal{W}_{ij} \right]}.
	\label{eq:M3C}
	\end{equation}

For comparison, the spins in an Ising system would be binary in orientation (i.e., $\theta \in \{0,\pi\}$). For any two ternary spins we have $\mathbf{s}_i \cdot \mathbf{s}_j = \cos ( \theta _i - \theta_j)\in $ $\left\{  1, -\frac{1}{2}\right\}$. This model is also known as the $q=3$ vector Potts model which has been studied in statistical mechanics and quite recently in the context of exciton-polariton condensates~\cite{Kalinin_PRL2018}. When $\theta_i \in [-\pi,\pi)$ then Eq.~\eqref{eqn:hamiltonain1} is just the standard XY model,
\begin{equation}
H_{XY} = -\sum_{ij}J_{ij} \cos{(\theta_i - \theta_j)}, \qquad \theta_i \in [-\pi,\pi).
\label{eqn:hamiltonain2}
\end{equation}
In the polariton network, the interacting quantities in question are nonlinear oscillators (the condensates) each characterized by a complex-valued number $\psi_i= \rho_i e^{i \theta_i}$. In this sense, the phasors of the condensates then take the role of interacting two-dimensional pseudospins $\mathbf{s}$ which dynamically experience a gradient descent towards the ground state of $H_{XY}$~\cite{berloff_realizing_2017}.

As pointed out by Frieze and Jerrum~\cite{Frieze_Alg1997}, a mapping exists between a max-$k$-cut problem and the ground state of a system of spins in the vertices of an equilateral simplex in $\mathbb{R}^{k-1}$. For the max-2-cut problem the corresponding simplex is a line with vertices $\theta = \{0,\pi\}$ which is the reason an Ising system ground state maps directly to the max-2-cut~\cite{barahona_computational_1982}. For the M3C problem the simplex is an equilateral triangle inscribed by the unit circle like in Eq.~\eqref{eqn:hamiltonain1} and shown in Fig.~\ref{fig:house}(a). For the max-4-cut the problem maps to the ground state of the Heisenberg spin system where the simplex is an equilateral tetrahedron inscribed by the unit sphere, and so on.

\subsection{The random hyperplane technique}

\begin{figure}[h!]
	\includegraphics[width=1.0\linewidth]{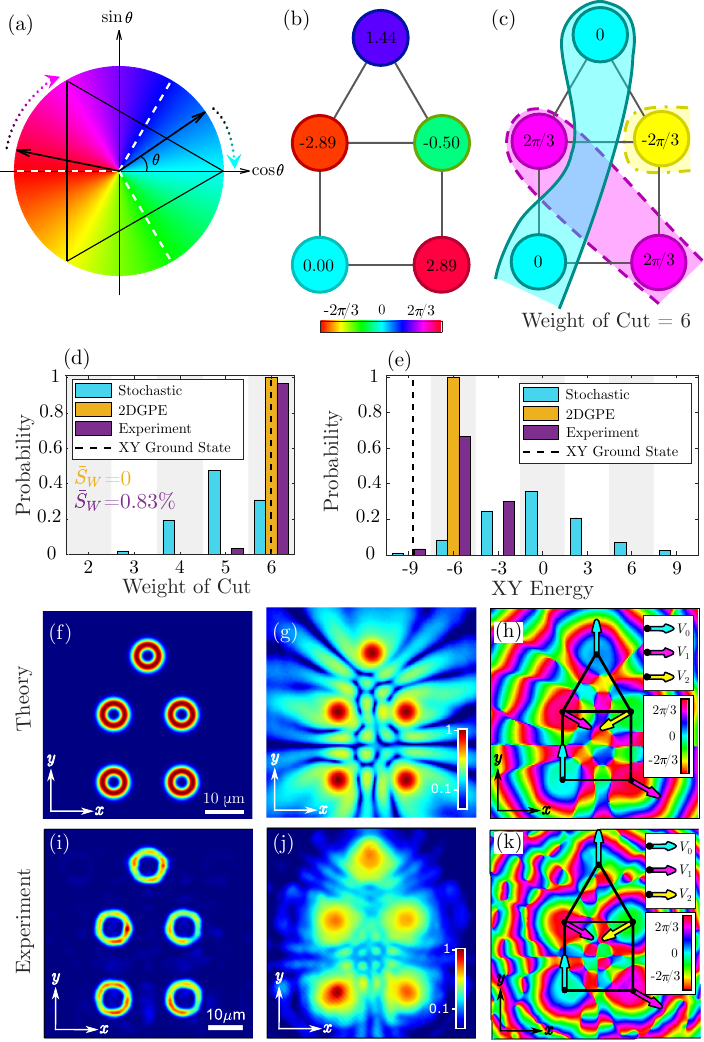}
	\caption{(a) Schematic of a chosen binning boundary (dashed white lines) projecting spins into the corners of the circumscribed triangle. (b) The XY ground state of the house configuration with numbers representing the angles (phases) $\theta_i$ of the vertices in radians; (c)  the ternary mapping of the phases in (b); (d, e) histograms of the M3C weights and XY energy respectively from the corresponding house graph of (blue) 1000 random samples of partitioning, (yellow) 2DGPE simulations and (purple) experiment. Dashed black lines show the XY ground state on spectra, not discretized to the histogram bins (white and gray background stripes). Weight of cuts in (d) represent the optimum cut for each data set sampled across 100 uniformally spaced binning boundaries (different rotations of the unit triangle) and the average error, $\bar{S}_W$, of the simulated and experimental cuts are shown in yellow and purple respectively. (f, i) Optical pump profile $P(\mathbf{r})$; (g, j) condensate density $|\Psi(\mathbf{r})|^2$, and (h, k) condensate phase map $\theta(\mathbf{r})$ from 2DGPE simulation and experiment, respective to each row, of optically trapped polariton condensates~\cite{harrison_synchronization_2020} in the AFM house configuration, with arrows in (h, k) showing the discretized XY phase into subsets $\mathcal{V}_{0,1,2}$.}
	\label{fig:house}
\end{figure}

To approach the ground state of the ternary Hamiltonian [Eq.~\eqref{eqn:hamiltonain1}] the spins of the XY Hamiltonian [Eq.~\eqref{eqn:hamiltonain2}] ground state configuration $\{\mathbf{s}_1, \mathbf{s}_2, \dots\}$, which can be approximately obtained using a polariton network~\cite{berloff_realizing_2017, Kalinin_SciRep2018}, are projected (or {\it binned}) onto their closest ternary counterpart corresponding to the vertices of the inscribed triangle [Fig.~\ref{fig:house}(a)]. The method of projecting continuous spin variables onto binary and ternary decision variables forms a semi-definite relaxation program which was studied in computer science by Goemans and Williamson~\cite{goemans_improved_1995, Goemans_JouCompSys2004} and referred to as the {\it random hyperplane technique}.
Here, we focus on the ternary problem where a binning boundary is depicted with dashed white lines in Fig. 1(a) like used in~\cite{Goemans_JouCompSys2004} which projects the continuous spins of our polariton network (see example black arrows) into their closest corner of the circumscribed triangle. An approximate solution of the ternary spin Hamiltonian is obtained through sampling many random orientations of the binning boundary with respect to the horizontal axis~\cite{Goemans_JouCompSys2004} (hence the name ``random'' hyperplane technique).
We point out that the minimization of the ternary [Eq.~\eqref{eqn:hamiltonain1}] and continuous [Eq.~\eqref{eqn:hamiltonain2}] XY system share the same relaxation procedure to a semi-definite program~\cite{goemans_improved_1995,zhang_complex_2006} underlining the common point of finding the ground states of the two systems.

\subsection{The house graph}

An example procedure is shown visually in Fig.~\ref{fig:house}(b) and~\ref{fig:house}(c) using the XY ground state of the anti-ferromagnetic (AFM) house graph, as previously studied in~\cite{gershenzon_exact_2020} using degenerate laser cavities. Here, AFM refers to $J_{ij}<0$ in Eqs.~\eqref{eqn:hamiltonain1} and \eqref{eqn:hamiltonain2} with preferential antiparallel spin alignement $\theta_i - \theta_j = \pi$ between two condensates/oscillators. The contrary, ferromagnetic (FM) alignment refers to preferential in-phase condensate/oscillator synchronization $\theta_{i}-\theta_j = 0$, corresponding to $J_{ij} > 0$ interactions. The AFM house graph consists of vertices [colored discs in Fig.~\ref{fig:house}(b)] arranged in the illustrated fashion with equally weighted AFM edges $J_{ij}=J<0$. For brevity, we will use $J=-1$ in dimensionless units. We first calculate the XY ground state using the conventional basin hopping optimization method~\cite{Wales_BasinHopping_1997} with the ground state angles (phases) $\theta_i$ given in radians inside the discs. An appropriately oriented binning boundary bins the angles into their ternary counterparts shown inside the discs in Fig.~\ref{fig:house}(c). We then apply this ternary XY outcome to the M3C problem through Eq.~\eqref{eq:M3C} which dictates that the partitions (and their cuts) follow the colored regions shown in Fig.~\ref{fig:house}(c). This is precisely the maximum cut of the M3C problem for the house graph, with weight $= 6$. It is worth noting that the minimum weight of a cut here is $=2$.

To illustrate that the maximum weight has been obtained, we plot in Fig.~\ref{fig:house}(d) the M3C weight (dashed black line) against the distribution of possible house graph weights obtained from 1000 random samples of $\theta_i$ tested against 100 random binning boundaries (cyan bars). The results show that stochastic sampling gives a wide spread in weights with a maximum around weight $=5$, underlining that even when many different boundaries are tested it does not guarantee good performance.

\section{Polariton dynamics}

To investigate the applicability of our method in polaritonic systems, we start by performing generalized two-dimensional Gross-Pitaevskii (2DGPE) simulations [see Eqs.~\eqref{eq:2DGPE}-\eqref{eq:ResA} in Appendix~\ref{app:B}] on five exciton-polariton condensates in the AFM house configuration using the technique described in~\cite{harrison_synchronization_2020}. For each steady-state realization, 100 random binning boundaries of the condensate phases are applied to find the M3C. We plot the resulting M3C weights and continuous XY energies (yellow bars) in Figs.~\ref{fig:house}(d,e), showing that simulations give a correct M3C weight $=6$ while coming close to the XY ground state as indicated by the dashed black line. We additionally plot the stochastically sampled distribution of $H_{XY}$ energies in cyan bars.

Next, we experimentally inject five exciton-polariton condensates in the house configuration in a semiconductor microcavity, similar to the methods detailed in~\cite{harrison_synchronization_2020}, and measure the output phase configuration of the interacting condensates using the techniques described in~\cite{Alyatkin_PRL2020} [see Appendix~\ref{app:C}]. We record 60 experimental realizations and calculate each time the XY energy of the condensate network, obtaining the distribution given by the purple bars in Fig.~\ref{fig:house}(e). Our observations confirm that interacting polariton condensates favor phase configurations towards small $H_{XY}$ energies as pointed out in~\cite{berloff_realizing_2017}, but might not necessarily reach the ground state and instead get stuck in local minima of the energy landscape, or converge into a nonstationary state where $\theta_i$ has no meaning. This explains why the theoretical and experimental distributions are maximal around $H_{XY} \approx -6$ instead of the ground state -8.7419. Nevertheless, the resulting M3C weights from the experiment shown in Fig.~\ref{fig:house}(d) (purple bars) indicate very good performance, implying that approximate solutions to the continuous variable problem of minimizing $H_{XY}$ can indeed give good results to the M3C. This important result underscores that coherent networks of dissipatively coupled oscillators, like polariton condensates, can potentially perform as heuristic solvers for the NP-hard M3C.

\begin{figure}[t]
	\centering
	\includegraphics[width=0.95\linewidth]{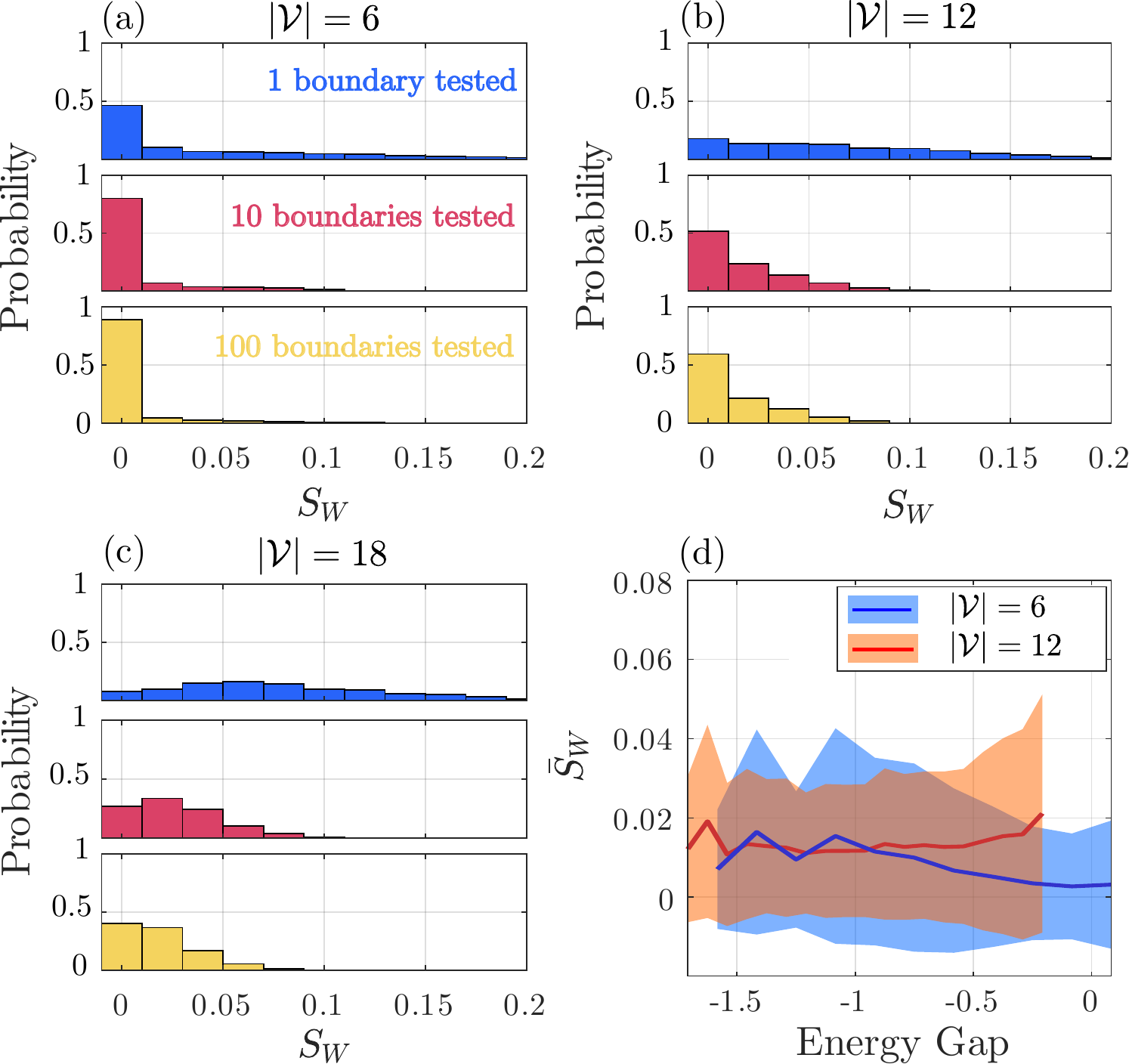}
	
	\caption{(a,b,c) Performance of $|\mathcal{V}|=6,12,18$ vertex graphs, respectively, showing an expected drop with graph size yet still maintaining peak probability around zero error ($S_W = 0$). The mean errors are $\bar{S}_W = 0.53 \%$, $1.38 \%$, and $1.86 \%$. With more random binning boundaries tested [different rotations of the triangle in Fig.~\ref{fig:house}(a)] the probability of finding the correct solution increases. (d) Mean error $\bar{S}_W$ (whole lines) plotted against the ground state energy gap, $\text{min}{(H_{XY})}- \text{min}{(H_T)}$, indicating weak dependence when the energies between the ternary and continuous spin systems are different. Shaded area denotes the standard deviation. Here, $\text{min}{(H_{XY})}$ is estimated using the Stuart Landau network (as opposed to e.g. the basin hopping method) for computational speediness, whereas $\text{min}{(H_T)}$ is found using brute force. Horizontal axis is given in units of $|\mathcal{V}| \sigma$ where $\sigma$ is the standard deviation of $\mathcal{W}_{ij}$.}
	\label{fig:benchmark}
\end{figure}

In Figs.~\ref{fig:house}(f,i) we show a real-space map of the non-resonant laser intensity used to excite the polariton condensates in simulation and experiment, respectively. The laser intensity is shaped to form five rings with central coordinates $\mathbf{r}_i = (x_i,y_i)$ which keep the condensates localized at the vertices of the house graph. The couplings $J_{ij}=J<0$ between the condensates which enter into Eqs.~\eqref{eqn:hamiltonain1} and~\eqref{eqn:hamiltonain2} are determined by choosing specific coordinates $\mathbf{r}_i$, as previously studied in Ref.~\cite{harrison_synchronization_2020}. In Figs.~\ref{fig:house}(g,h) and \ref{fig:house}(j,k) we show the density $|\Psi(\mathbf{r})|^2$ and the phase $\theta(\mathbf{r}) =\arg{(\Psi(\mathbf{r}))}$ of the condensate wavefunction $\Psi(\mathbf{r})$ from simulation and experiment respectively, where Fig.~\ref{fig:house}(j) is averaged over many condensate realizations [see Fig.~\ref{fig:exp}(c) for the single shot real-space realization of the house graph PL]. The clear interference pattern observed in both experiment [Fig.~\ref{fig:house}(j)] and simulation [Fig.~\ref{fig:house}(g)] implies phase synchronization between condensates. With the condensates synchronized, we can extract their phases at the location of the vertices through interferometric techniques~\cite{Alyatkin_PRL2020} such that $\theta(\mathbf{r}_i) = \theta_i$. The presented phase maps in Figs.~\ref{fig:house}(h) and~\ref{fig:house}(k) both give a ternary phase configuration (overlaid) that matches the M3C shown in Fig.~\ref{fig:house}(c).

\subsection{Benchmarking phase-and-amplitude oscillators}

The results shown in Fig.~\ref{fig:house} underline the promise of applying dissipative oscillatory systems to solve the M3C problem. But in order to gain a better understanding on the quality of the method we look into its statistics by testing many different graph configurations $\mathcal{G}$ of randomly chosen weights $\mathcal{W}_{ij}$. In Fig.~\ref{fig:benchmark} we test the performance of our method to solve the M3C by simulating a network of dissipatively coupled phase-and-amplitude (Stuart-Landau) oscillators [see Eq.~\eqref{eq:0DGPE}] which form a very general setting of coupled nonlinear oscillators. We compare the weight obtained from the oscillator network $W_\text{SL}$ against the correct maximum/minimum weight $W_\text{BF}^\text{max(min)}$ belonging to a 3-cut in the graph $\mathcal{G}$, found using a brute force method. We define the normalized error as, 
\begin{equation}
S_W = \frac{W_\text{BF}^\text{max} - W_\text{SL}}{W_\text{BF}^\text{max} - W_\text{BF}^\text{min}}.
\end{equation}
With this metric $S_W = 0$ means that the system has found the best cut whereas $S_W=1$ is the worst cut. We use 10000 dense random configurations $\mathcal{W}_{ij}$ (normally distributed with zero mean) and numerically solve the Stuart-Landau network dynamics from stochastic initial conditions for each configuration to obtain the steady (synchronized) state, corresponding to fixed points of $\psi_n^* \psi_m$. Increasing the number of random binning boundaries tested results in increased performance which can be intuitively understood from the fact that while Eq.~\eqref{eqn:hamiltonain2} is independent of global rotation of the spins $\theta_i \to \theta_i + \phi$ the procedure of binning the XY spins to evaluate Eq.~\eqref{eqn:hamiltonain1} is not. Therefore several different orientations of the binning boundary should be attempted in order to obtain the best value to the M3C. The results show very good performance with mean error of $\bar{S}_W = 0.53 \%, 1.38 \%$, and $1.86 \%$ for graphs of sizes $|\mathcal{V}| = 6, \ 12$, and $18$, well below today's best error guarantee of 16.40\% in~\cite{Goemans_JouCompSys2004} and 19.98\% in \cite{Frieze_Alg1997}. 

We additionally investigate whether the performance of the system depends on the energy gap between the ground states of the ternary spin system and continuous $XY$ spin system, i.e. $\text{min}{(H_{XY})}- \text{min}{(H_T)}$. Results in Fig.~\ref{fig:benchmark}(d) show the mean error for different energy gaps, indicating good performance with no significant dependence on the ground state energy gap between the two Hamiltonians.

\begin{figure}[t]
	\centering
	\includegraphics[width=1.0\linewidth]{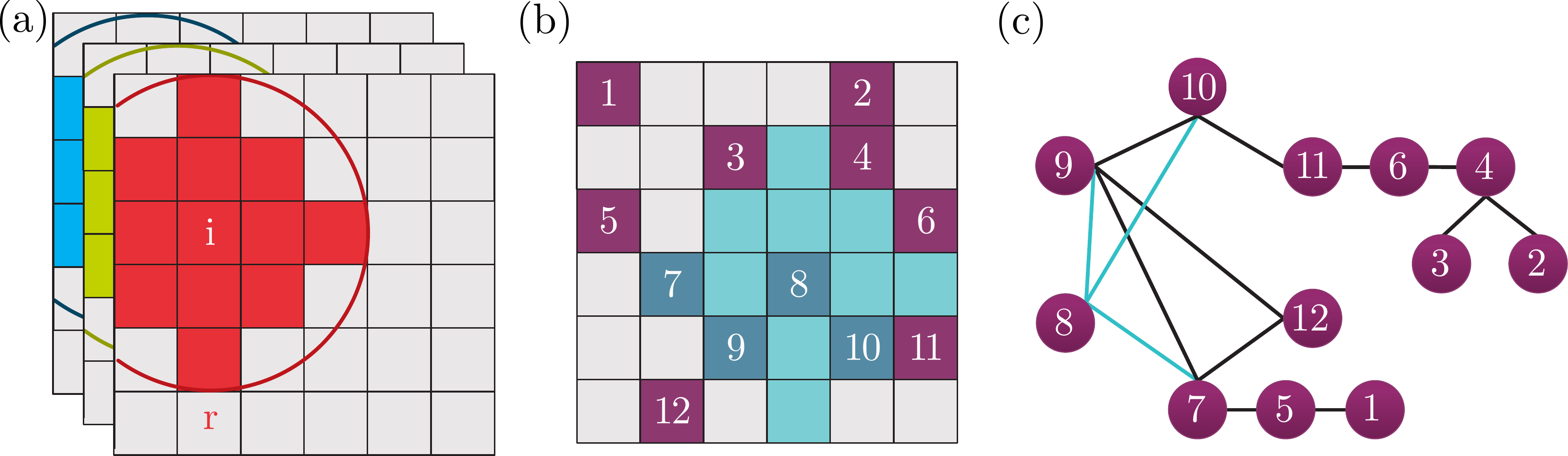}
	
	\caption{(a) Each small square represents a pixel in a colored image with $i$ labelled and the surrounding patch colored according to RGB color layer. Radius $r$ shows the maximum distance of pixel connectivity where only pixels within the ring have an edge weight connecting $\mathcal{V}_i$ and $\mathcal{V}_j$, otherwise $\mathcal{W}_{ij}=0$; (b) $m=12$ random pixels are selected and transformed into a graph (c).}
	\label{fig:imSeg}
	
\end{figure}

\section{Image-Segmentation}
The M3C problem can be used to segment an image into 3 objects or regions~\cite{felzenszwalb_efficient_2004}. When formulating a graph from an image, the vertices and edges represent pixels and their relative similarity respectively. When selecting vertices, all pixels or a smaller sample can be included, with connectivity to neighbours within a given radius $r$, as shown in Fig.~\ref{fig:imSeg}(a). To define the edge weights $\mathcal{W}_{ij}$, both local and global properties between two sampled pixels $i$ and $j$ such as color, brightness, texture, and spatial proximity~\cite{felzenszwalb_efficient_2004, rouco_automatic_2016, martin_database_2001} can be used through variety of weight estimating methods [see Eqs.~(\ref{eq:M1}-\ref{eq:M5}) in Appendix~\ref{app:D}]~\cite{boykov_graph_2006,wang_graph-based_2013} to form a graph with vertices representing the sampled pixels [Fig.~\ref{fig:imSeg}(b,c)].
\begin{figure}[b]
	
	\includegraphics[width=1.0\linewidth]{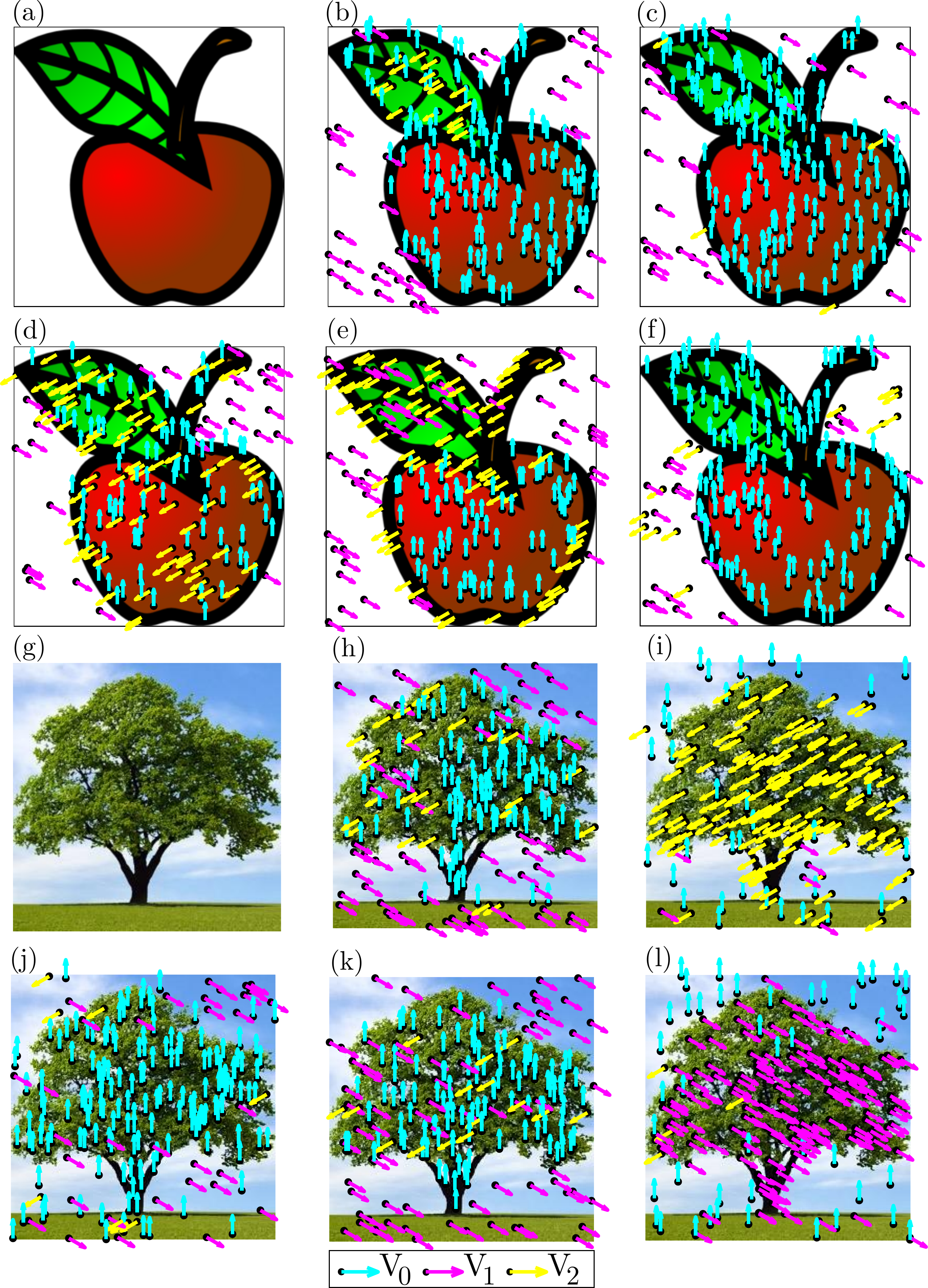}
	
	\caption{Images of (a) an apple and (g) a tree, with image-segmentation results (b-f, h-l) using the Stuart-Landau network for methods Eqs.~\eqref{eq:M1}-\eqref{eq:M5} respectively, with $m=200$, $r=400$, $q=0.1$ for (b-f) and $q=0.2$ for (h-l) [see Appendix~\ref{app:D}]. The cyan, magenta and yellow arrows represent the subsets $\mathcal{V}_{0,1,2}$ respectively.}
	\label{fig:tree}
\end{figure}

We show 3-way image-segmentation for pictures of an apple and a tree by solving the M3C problem for each graph using the same procedure as described in Figs.~\ref{fig:house}-\ref{fig:benchmark} with the Stuart-Landau network [see Eq.~\eqref{eq:0DGPE}]. In Fig.~\ref{fig:tree} we show the partitioned spins overlaid on example images of an apple and a tree for five different methods of estimating $\mathcal{W}_{ij}$ that all successfully locate objects within each image. Methods which combine both local (near neighbour) pixel and global properties [Figs.~\ref{fig:tree}(c,f,i,l)] best locate a single object and a background by predominantly cutting the graph into just two subsets
. This is a result of the similar-valued edge weights between pixels within the red, black and green regions. However, the
methods that only consider local pixel values [Figs.~\ref{fig:tree}(b,e,h,k)] segment the images into multiple same-spin regimes. This is seen more clearly for the simpler apple image, where the local  methods are able to locate the white background, black outline, red body and green leaf.
As there are four main block colors making up the apple image but only three subsets for the segmentation, a single subset is representing multiple objects, such as $\mathbf{\mathcal{V}_0}$ in Fig.~\ref{fig:tree}(b) representing both the red and black regions of the apple.
By considering only global image properties [Figs.~\ref{fig:tree}(d,j)], the images also segment into object and background by locating the dominant colors of the image, though some objects are located by multiple subsets.
Again, this is an artefact due to having only three subsets to segment a four object image with. In Fig.~\ref{fig:tree}(d) for example, the red of the apple is represented by both $\mathcal{V}_0$ and $\mathcal{V}_2$ which indicates that the steady state phases of these oscillators fell about the binning boundary between these two segmentations. Through adjusting the segmentation parameters and choosing a specific method, image segmentation using dissipative coupled oscillators can be achieved to match a wide range of segmentation requirements.

\begin{figure}[t]
	\centering
	\includegraphics[width=0.97\linewidth]{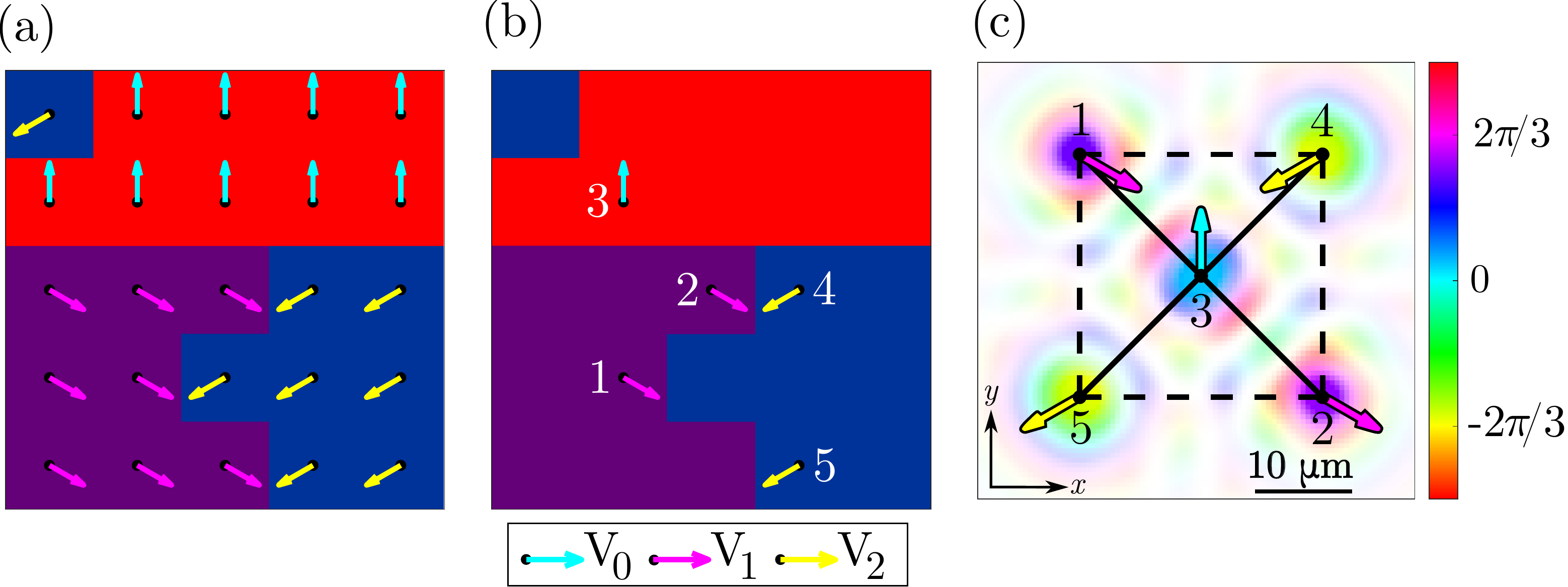}
	
	\caption{Simple colored image (a,b) showing image-segmentation using the Stuart-Landau network for the phase-discretized XY ground state of the graphs following Eq.~\eqref{eq:M4}, with $r=4$, $q=0.01$ for (a) $m=25$ and (b) $m=5$. (c) Shows the condensate phase map $\arg{(\Psi(\mathbf{r}))}$ obtained from 2DGPE simulation solving the set of pixels given by (b) with transparency proportional to the polariton density $|\Psi(\mathbf{r})|^2$. Projecting the phase of each condensate into its ternary counterpart is given by the cyan, magenta and yellow arrows, representing the subsets $\mathcal{V}_{0,1,2}$ respectively. Solid and dashed black lines represent $\mathcal{W}_{ij}=1.0$ and $\mathcal{W}_{ij}=0.8$ (rounded to 1 decimal place) respectively.}
	\label{fig:simple}
	
\end{figure}

We also consider a simpler colored image with 25 pixels in Fig.~\ref{fig:simple} to demonstrate image-segmentation using a planar graph. We first find the image-segmentation again using the Stuart-Landau network, with the all-to-all coupling between 25 pixels [see Fig.~\ref{fig:simple}(a)], and a random sample of 5 pixels [see Fig.~\ref{fig:benchmark}(b)], which correctly locates the different block colors in the image. As the sampled pixels are sparsely connected in the latter approach, the correct image-segmentation can be solved using a small planar graph of polariton condensates like in Fig.~\ref{fig:house} which, through optimization of the M3C, converge to a phase map (Fig.~\ref{fig:simple}(c)) representing the correct image-segmentation.

\section{Constrained-Via-Minimization}
Here we discuss the application of the M3C for constrained-via-minimization (CVM) in circuit design using coherent networks. In order to optimize space used in commercial applications, complex circuits are often split over multiple layers of a circuit board. This is achieved by drilling holes, known as ``vias'', that are lined with a conductive coating, allowing tracks to connect between multiple layers. These additional vias increase production time, complexity, and cost, making it desirable to minimize their number.

\begin{figure}
	\centering
	\includegraphics[width=1\linewidth]{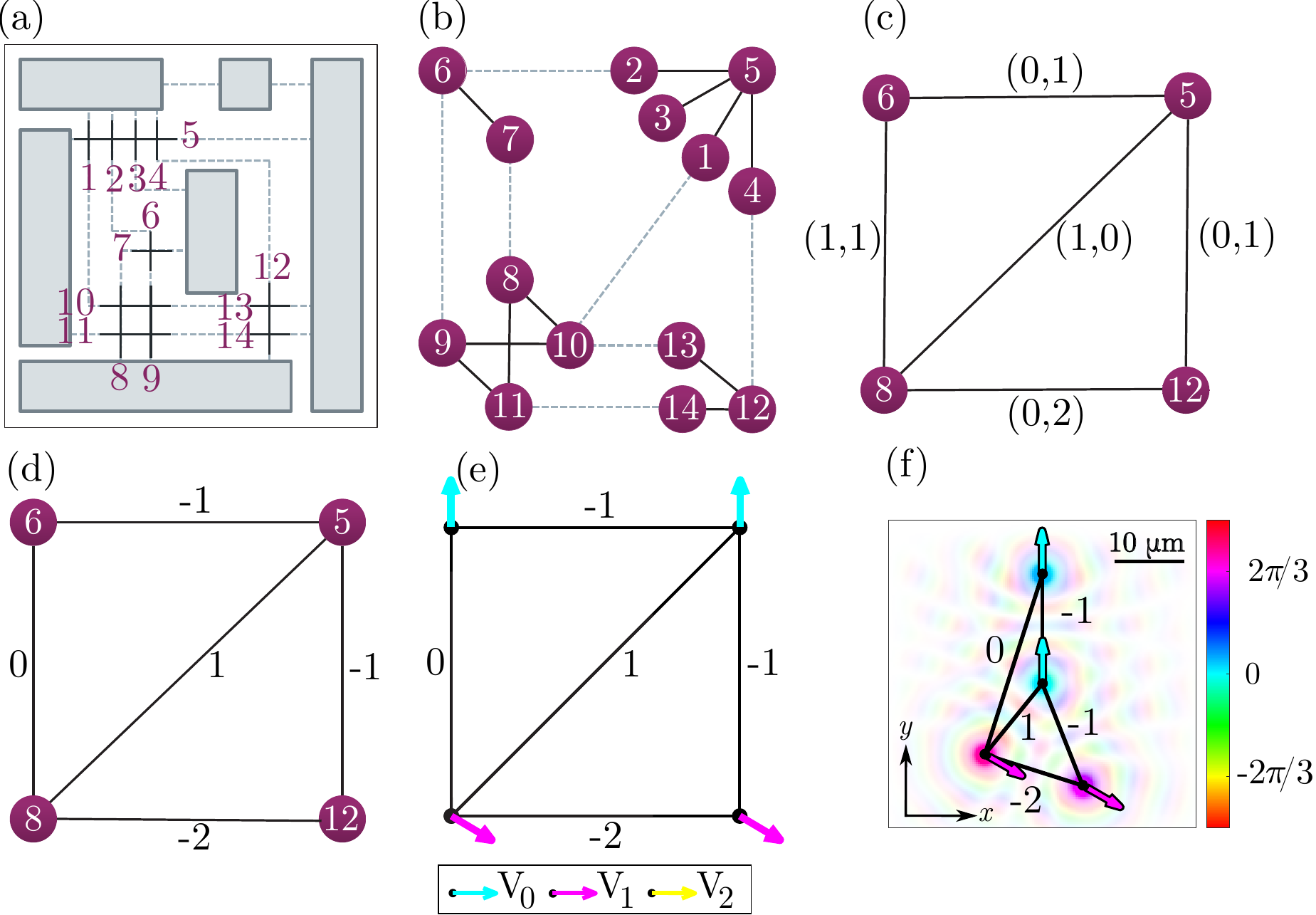}

	\caption{(a) A circuit with routing between pre-placed cells (gray areas), with critical segments of tracks numbered and shown in solid black line and free segments showed in gray dashes; (b) the layout graph of circuit (a) with critical edges shown in solid black and continuation edges in gray dashes; (c) the reduced layout graph of (b) with edges labelled ($\alpha_{ij},\beta_{ij}$); (d) the reduced layout graph of (b) with edge weights $w_{ij}=\alpha_{ij}-\beta_{ij}$, and (e,f) ground state discretized XY phase for the M3C of graph (d), solved using the Stuart-Landau network and 2DGPEs respectively, with black numbers representing the graph edge weights $w_{ij}$. The cyan, magenta and yellow arrows represent the subsets $\mathcal{V}_{0,1,2}$ respectively and in (f) the transparency is proportional to the polariton density $|\Psi(\mathbf{r})|^2$.}
	\label{fig:cvm}
\end{figure}

For CVM, all cells (gray areas in Fig.~\ref{fig:cvm}(a)) are pre-placed and vertical and horizontal tracks are routed with the assumption that all pin connections are bipartite (i.e., a single track connects two pins), but layer assignment is not yet performed. Segments of track which overlap are labelled as critical segments (solid lines) which cannot be on the same layer of circuit board. Free segments (dashed lines) of track have no overlap with other tracks and these are the regions in which vias can be placed. We demonstrate CVM through reducing the task to a M3C problem~[34], which we then solve through simulation on a network of polariton condensates. We demonstrate CVM to a maximum of three layers of circuit board through reducing the task to a M3C problem. It is worth noting that this exact CVM problem is tackled in the work of~\cite{barahona_application_1988} with the max-2-cut problem, so we know that the minimum number of vias for the circuit in Fig.~\ref{fig:cvm}(a) requires just two circuit board layers, such that the solution to the max-3-cut problem will only incorporate two out of the three possible spin subsets.

For a circuit [see Fig.~\ref{fig:cvm}(a)], we define the layout graph $\mathcal{G} = (\mathcal{V},\mathcal{E})$. Each critical segment is represented by a vertex in set $\mathcal{V}$, where pairs of vertices are connected by either a conflict edge $\mathcal{A}$ (when a pair of critical segments cross paths) or by a continuation edge $\mathcal{B}$ (when a pair of critical segments are connected by a free segment), such that $\mathcal{E}=\mathcal{A}\cup\mathcal{B}$. The layout graph of the example circuit is shown in Fig.~\ref{fig:cvm}(b).
The reduced layout graph $\mathcal{R}=(\mathcal{S},\mathcal{T})$ arbitrarily selects a vertex $v_i$ in $\mathcal{V}_i$ to represent critical region $i$, such that $\mathcal{S} = \{v_i, \dots,v_z\}$. $\mathcal{T}$ contains the edges linking $v_i$ and $v_j$ for $i \neq j$, if and only if $\mathcal{G}$ contains a continuous edge connecting some vertex in $\mathcal{V}_i$ to some vertex in $\mathcal{V}_j$. As continuation edges do not cross, $\mathcal{R}$ is a planar graph. The edges of $\mathcal{R}$ in Fig.~\ref{fig:cvm}(c) have weights ($\alpha_{ij}$, $\beta_{ij}$), as contained in $\mathcal{T}$, such that:

\begin{center}
	\begin{tabular}{p{0.05\textwidth}p{0.35\textwidth}}
		
		$\alpha_{ij} =$& Sum of free segments between $\mathcal{V}_i$ and $\mathcal{V}_j$ connecting two critical segments with different orientations ; \\
		
		$\beta_{ij} =$& Sum of free segments between $\mathcal{V}_i$ and $\mathcal{V}_j$ connecting two critical segments with the same orientation.
	\end{tabular}
\end{center}

The partition of this graph into $\mathcal{S}_n$, $n=0,1,2$, corresponds to the assignment of each critical region to 3-layer $n$. For such a partition, the number of vias required is
\begin{equation}
\text{VIA}(\mathcal{S}_n) = \sum_{v_iv_j\in \mathcal{T}_n}\alpha_{ij} + \sum_{v_iv_j\in \delta \mathcal{T}}\beta_{ij}
\end{equation}
where edges $\mathcal{T}_n$ connect critical regions assigned the same layer and $\delta \mathcal{T}$ connect critical regions assigned different layers. By defining $A = \sum_{v_iv_j\in \mathcal{T}}\alpha_{ij}$, then
\begin{equation}
\text{VIA}(\mathcal{S}_n) - A = \sum_{v_iv_j\in \delta \mathcal{T}}(\beta_{ij}-\alpha_{ij}).
\end{equation}
As $A$ is invariant on the layer assignment, and by redefining the edge weights $\mathcal{W}_{ij}=\alpha_{ij} - \beta_{ij}$ as in Fig.~\ref{fig:cvm}(d), the problem is reduced to a M3C problem,
\begin{equation}
\max \sum_{v_iv_j \in \delta \mathcal{T}} \mathcal{W}_{ij}.
\end{equation}
The phase-discretized XY solution coming from the polariton network for our example reduced layout graph [see Figs.~\ref{fig:cvm}(e,f)] partitions the vertices into just two subsets, showing that the minimum via configurations only requires two layers of circuit board. In this example, the minimized number of vias is 2, as $A - \sum_{v_iv_j \in \delta \mathcal{T}} \mathcal{W}_{ij} = 2 - 0$ and thus a correct solution has been found by the polariton network.

\section{Conclusions}
We have investigated both theoretically and experimentally the potential of using nonlinear optical oscillatory networks, specifically exciton-polariton condensates, in approximating the solutions to the NP-hard max-3-cut optimization problem. Our study is motivated by recent works showing that networks of exciton-polariton condensates undergo a gradient descent towards equalibria of synchronized states with phasor configurations that correlate with the ground state of the XY Hamiltonian~\cite{berloff_realizing_2017,Kalinin_SciRep2018,Kalinin_PRL2018, harrison_synchronization_2020}. We exploit this dynamical feature to approximate the max-3-cut through two methods: First, we apply a semi-definite relaxation program by Goemans and Williamson known as a random hyperplane method~\cite{goemans_improved_1995, Goemans_JouCompSys2004} which projects (bins) the condensate phasors into ternary decision variables. Second, we use the direct mapping between the $q=3$ vector Potts model (which we refer to as the ``ternary XY model'') and max-3-cut, as originally presented by Frieze and Jerrum~\cite{Frieze_Alg1997}, which can then be approximated using the previously obtained ternary decision variables. Our study provides experimental evidence that polariton condensate networks can potentially serve as optical annealers for the max-3-cut, and opens perspectives on the role of coherent nonlinear optical networks as fast approximate analogue solvers for complex combinatorial problems. While interactions between polariton condensates are inherently planar work has begin to address the possibility of all-to-all connectivity~\cite{Sigurdsson_ACSPho2019, Kalinin_NanoPho2020, Harrison_arxiv2021}. We have also studied the more complex applications of image-segmentation and CVM in circuit design which we heuristically solve using our proposed method and simulations on both Stuart-Landau oscillator networks and polariton condensate networks.

Our work goes beyond previously studied Ising machines to solve the max-2-cut problem~\cite{inagaki_coherent_2016} by exploiting the continuous phase degree-of-freedom in oscillatory systems. Their natural tendency in minimizing the XY model can be applied to solve the M3C problem through Goemans and Williamson inspired semi-definite program, where we further explore the projection of phases to 2 and 4 bins to solve the max-2-cut and max-4-cut in Appendix~\ref{app:E}. This technique can be applied to any dissipative oscillatory system such as laser networks and photonic condensates, but in this study we have taken steps towards realizing a liquid light machine of interacting polariton condensates as a practical coherent network computational device by exploiting the ultrafast temporal dynamics, parallel interactive nature, and continuous degree of freedom that can now be readily accessed in state-of-the-art experiments in polariton lattices~\cite{Alyatkin_PRL2020,topfer_engineering_2021}.

\section*{Acknowledgements} 
S.L.H., H.S., J.D.T. and P.G.L. acknowledge the support of the UK’s Engineering and Physical Sciences Research Council (grant EP/M025330/1 on Hybrid Polaritonics), and S.A. acknowledges the funding of the Russian Foundation for Basic Research (RFBR) within the joint RFBR and CNR project No. 20-52-7816. H.S. and P.G.L. also acknowledge the European Union’s Horizon 2020 program, through a FET Open research and innovation action under the grant agreement No. 899141 (PoLLoC). S.L.H. acknowledges the use of the IRIDIS High Performance Computing Facility, and associated support services at the University of Southampton, in the completion of this work.

\section*{Data Availability}
The data presented in this manuscript can be accessed on the University of Southampton data repository by following: \href{https://doi.org/10.5258/SOTON/D2104}{doi.org/10.5258/SOTON/D2104}.
\appendix

\section{Polariton Theory}\label{app:B}
The ground state of the XY model is obtained numerically through previously studied methods using the generalized GPE equation describing the dynamics of interacting polariton condensates~\cite{berloff_realizing_2017}. The laser driven microcavity system is simulated close, but above, the condensation threshold where the interacting condensates (corresponding to their respective laser spots) are found to spontaneously self-organize into a phase configuration which maximizes the particle number of the condensate which can be regarded as minimization of an effective XY model~\cite{lagoudakis_polariton_2017}. The polariton condensate wavefunction, $\Psi(\mathbf{r},t)$, is described by the generalized GPE coupled to a reservoir of excitons $n(\mathbf{r},t)$ which experience bosonic stimulated scattering into the condensates~\cite{Wouters_PRL2007}.
\begin{multline}
i \frac{\partial \Psi}{\partial t} = \bigg[-\frac{\hbar \nabla^2}{2m} + G \left(n + \frac{P(\mathbf{r})}{W} \right) \\+  \alpha |\Psi|^2 + \frac{i }{2} \left( R n - \gamma \right) \bigg] \Psi,
\label{eq:2DGPE} 
\end{multline}

\begin{align}
\frac{ \partial n}{\partial t} & = - \left( \Gamma + R |\Psi|^2 \right) n + P(\mathbf{r}). \label{eq:ResA}
\end{align}

Here, $m$ is the effective mass of a polariton in the lower dispersion branch, $\alpha$ is the interaction strength of two polaritons in the condensate, $G$ is the polariton-reservoir interaction strength, $R$ is the rate of stimulated scattering of polaritons into the condensate from the reservoir, $\gamma$ is the polariton decay rate, $\Gamma$ is the decay rate of the the reservoir excitons, $W$ quantifies the population ratio between low-momentum excitons that scatter into the condensate and those which reside at higher momenta (so called inactive excitons), and $P(\mathbf{r})$ is the non-resonant continuous wave (CW) pump profile given by
\begin{equation} \label{eq.Pr}
P(\mathbf{r}) = P_0 \sum_i p(\mathbf{r}-\mathbf{r}_i).
\end{equation}
Here, $P_0$ denotes the laser power density and the function $p(\mathbf{r})$ corresponds to the 2D annular shaped profile of a single laser incident onto the microcavity plane and the coordinates $\mathbf{r}_i$ are the locations of the vertices in the polariton graph. The annular shaped pump profiles optically trap each condensate, yet allow for coherent transport of particles between nearest neighbours, consistent with those described in~\cite{harrison_synchronization_2020}.

In the 2DGPE simulations, the parameters are taken such that the polariton mass and lifetime are based on the properties of a laboratory InGaAs microcavity sample: $m = 0.28$ meV ps$^2$ $\upmu$m$^{-2}$ and $\gamma = \frac{1}{5.5}$ ps$^{-1}$. We choose values of interaction strengths typical of InGaAs based systems: $\hbar \alpha = 7$ $\upmu$eV  $\upmu$m$^2$, $G = 10 \alpha$. The reservoir decay rate is taken comparable to the condensate decay rate $\Gamma = \gamma$ due to fast thermalization to the exciton background. The final two parameters are then found by fitting to experimental results where we use the values $\hbar R = 98.9$ $\upmu$eV $\upmu$m$^{2}$, and $W = 0.035$ ps$^{-1}$.

We also consider a more simple and general model which does not depend on complicated spatial degrees of freedom. It describes dissipative coupling between nonlinear oscillators $\psi_n(t)$ and is known as a Stuart-Landau network,
\begin{equation}
\frac{d \psi_n}{dt} = \left[ P  -   |\psi_n|^2 \right]\psi_n + \sum_m J_{nm} \psi_m.
\label{eq:0DGPE}
\end{equation}
Here, $P$ denotes the ``gain'' of each oscillator and $J_{nm}$ is the coupling strength. Such networks have shown good performance in minimizing the XY Hamiltonian~\cite{Kalinin_SciRep2018} and share similarities to adiabatic bifurcation networks~\cite{Goto_ScieAdv2019}. Our simulations always start from random initial conditions and once the oscillators $\psi_n$ have converged to a steady state for a given $P$ we extract their phases $\theta_n = \text{arg}{(\psi_n)}$ to obtain the approximate ground state energy of Eq.~\eqref{eqn:hamiltonain2}. We perform numerical integration of Eqs.~\eqref{eq:2DGPE}-\eqref{eq:ResA} and Eq.~\eqref{eq:0DGPE} in time using a linear multistep method. We point out that, to the best of our knowledge, there is no analytical estimate on the ``performance guarantee'' of the simulated Stuart-Landau network in finding the XY ground state and therefore the performance guarantee of finding the M3C cannot be ascertained at the current stage except through numerical methods.

In the Stuart-Landau network simulations we increase $P$ linearly in time from $P(t=0) = -\lambda_\text{max}$ to $P(t=T) = \lambda_\text{max}$ where $T$ is the total integration time and $\lambda_\text{max}>0$ is the largest eigenvalue of the coupling matrix $\mathbf{J} = \{ J_{nm} \} \in \mathbb{R}^{N \times N}$. This physically replicates a slow ramp-up in laser power beyond the polariton condensation threshold resulting in measurable photoluminescence from the system. We note that in the terms of amplitude oscillator models, the condensation threshold is a bifurcation point marking the departure of the condensate (the oscillator) from the stable $|\psi_n|=0$ solution. In Fig.~\ref{fig:m2c_vs_m3c} we set $P=0$ to underline that our results do not critically depend on slow pump rise times.

\section{Experiment}\label{app:C}

\begin{figure}[h]
	\centering
	\includegraphics[width=1.0\linewidth]{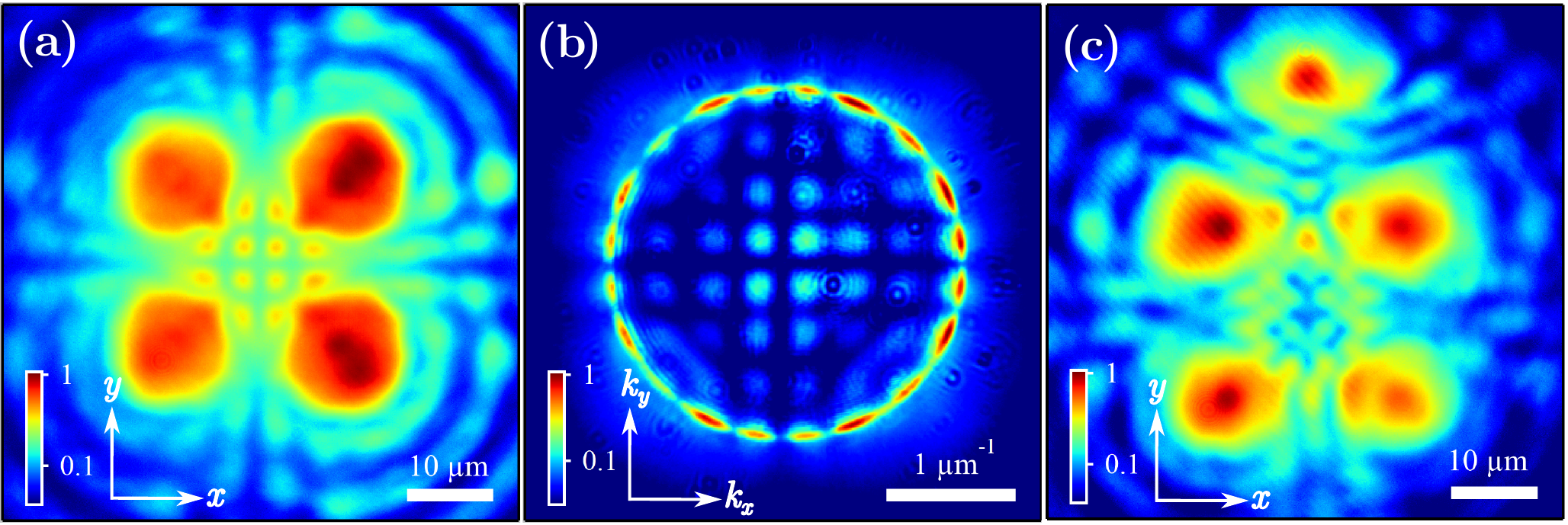}
	\caption{Time integrated (a) real-space and (b) reciprocal-space polariton PL for the square cell of trapped condensates sustaining AFM coupling. (c) Single shot realization of the polariton PL in real space for the AFM house graph.}
	\label{fig:exp}
\end{figure}

The GaAs-based microcavity is cooled down to 4 K in a closed-cycle helium crysotation and pumped non-resonantly with a CW single-mode laser at a negative exciton-photon detuning of -4.2 meV. In order to shape the excitation profile we utilize a programmable reflective phase-only spatial light modulator. The laser pump pattern is focused onto the sample with a 50x microscope objective of NA=0.42. The separation distance between pumping rings (20.5 $\upmu$m) and their diameter (8.1 $\upmu$m) is chosen such that any pair of the nearest trapped polariton condensates in a square geometry demonstrates AFM coupling [see Figs.~\ref{fig:exp}(a,b)]. The excitation geometry defines the range of the excitation intensity per single pumping ring, where the trapped condensates maintain AFM coupling. Therefore, for the house configuration the excitation
power is proportionally increased to ensure AFM coupling between nearest condensates. The results presented in Fig.~\ref{fig:house} correspond to excitation conditions supporting only single energy (i.e. stationary) states above condensation threshold. Time-averaged measurements of real space polariton photoluminescence are performed under CW excitation, acousto-optically modulated in time with square pulses at a frequency of 5 kHz and duty cycle of 1$\%$. To implement the relative phase readout between the nodes in the house configuration we utilize a homodyne interferometric technique~\cite{Alyatkin_PRL2020}. Each reconstructed phase map is extracted from a single-shot measurement in a Mach-Zenhder interferometer. The excitation pulsewidth in all single-shot measurements is 100 $\upmu$s.

The results of the single shot measurements are shown in Figs.~\ref{fig:house}(d,e), on the purple histograms of the M3C weights and XY energy respectively. Here, in Fig.~\ref{fig:exp}(c) we show the single shot polariton PL in real space for the house graph that gives the best result in minimizing the XY energy of the system. The corresponding phase map for this realization is shown in Fig.~\ref{fig:house}(k).


\section{Image-Segmentation}\label{app:D}
For a colored image, we randomly sample $m$ pixels. For each color layer, we consider each sampled pixel $i$ and define a local patch encompassing it and its (up to) eight-way nearest neighbours, as depicted in Fig.~\ref{fig:imSeg}(a). We define the sum of all $N$ pixels in the patch as,
\begin{equation}
p_i = \sum_{i=1}^{N}i,
\end{equation}
where $p_i$ is calculated for all pixels in the image and are normalized over each color layer
~\cite{wang_graph-based_2013}.
In addition to these local weights, we assign a global color weight, $c_i = M_A/M$, to each pixel according to its color frequency in the image, where $A$ is the color of pixel $i$, $M_A$ is the number of pixels with color $A$, and $M$ is the total number of pixels. We define the edge weight connecting pixels $i$ and $j$ as $\mathcal{W}_{ij}$, which is equal to 0 if the pixel separation is greater than $r$. Otherwise, we define five methods for enumerating $\mathcal{W}_{ij}$ between two sampled pixels based on a variety of techniques described in literature~\cite{felzenszwalb_efficient_2004, wang_graph-based_2013,peng_survey_2013,boykov_graph_2006} that utilize the difference between their pixel values, patch values and global weighting of their colors, as well as the difference between the product of their global color weighting with the patch or pixel values,
\begin{equation}
\hspace{0.055\textwidth}\text{Method 1:}\hspace{0.01\textwidth}\mathcal{W}_{ij} = \exp \bigg( \frac{|p_{i}-p_{j}|^{q}}{\sigma} \bigg);
\label{eq:M1}
\end{equation}

\begin{equation}
\hspace{0.08\textwidth}\text{Method 2:}\hspace{0.01\textwidth}\mathcal{W}_{ij} = \exp \bigg( \frac{|c_ip_{i}-c_jp_{j}|^{q}}{\sigma} \bigg);
\label{eq:M2}
\end{equation}

\begin{equation}
\hspace{0.05\textwidth}\text{Method 3:}\hspace{0.01\textwidth}\mathcal{W}_{ij} = \exp \bigg( \frac{|c_{i}-c_{j}|^{q}}{\sigma} \bigg);
\label{eq:M3}
\end{equation}

\begin{equation}
\text{Method 4:}\hspace{0.01\textwidth}\mathcal{W}_{ij} = \exp \bigg( \frac{|i-j|^{q}}{\sigma} \bigg);
\label{eq:M4}
\end{equation}

\begin{equation}
\hspace{0.065\textwidth}\text{Method 5:}\hspace{0.01\textwidth}\mathcal{W}_{ij} = \exp \bigg( \frac{|ic_{i}-jc_{j}|^{q}}{\sigma} \bigg).
\label{eq:M5}
\end{equation}
Here, $\sigma$ is the standard deviation in brightness across each patch, $q$ is a free parameter and $\mathcal{W}_{ij}$ is averaged over each color layer. To find the results of the image-segmentation, we find the M3C of the graph $\mathcal{G} = (\mathcal{V},\mathcal{W})$, with $|\mathcal{V}| = m$ vertices representing the sampled pixels with weights $\mathcal{W}_{ij}$ connecting vertices $i$ and $j$. We will show that the partition of the continuous phases in a polariton condensate graph (just like demonstrated in Fig.~\ref{fig:house} for the house graph) into a ternary phase configuration, acting as decision variables for the M3C, can segment different objects within an image.

\section{Other maximum cuts}\label{app:E}
It is worth noting that the continuous-phase spins of the XY model can also be easily projected onto binary angles 0 and $\pi$ (i.e. to 1-dimensional spins) which leads to an Ising energy function whose ground state can be mapped to the max-2-cut (M2C) problem~\cite{barahona_application_1988}. Such a binary projection strategy has been previously explored with coherent photonic Ising machines~\cite{marandi_network_2014, mcmahon_fully_2016, inagaki_coherent_2016, Inagaki_NatPho2016, Kyriienko_PRB2019, Pierangeli_PRL2019, RoquesCarmes_NatCom2020}. However, to the best of our knowledge, continuous phase photonic machines have not been used to explore the mapping between the ground state of the ternary spin Hamiltonian and the M3C problem (i.e., projection onto [0, 2$\pi$/3, 4$\pi$/3] spins).
	
\begin{figure}[t]
	\centering
	\includegraphics[width=1\linewidth]{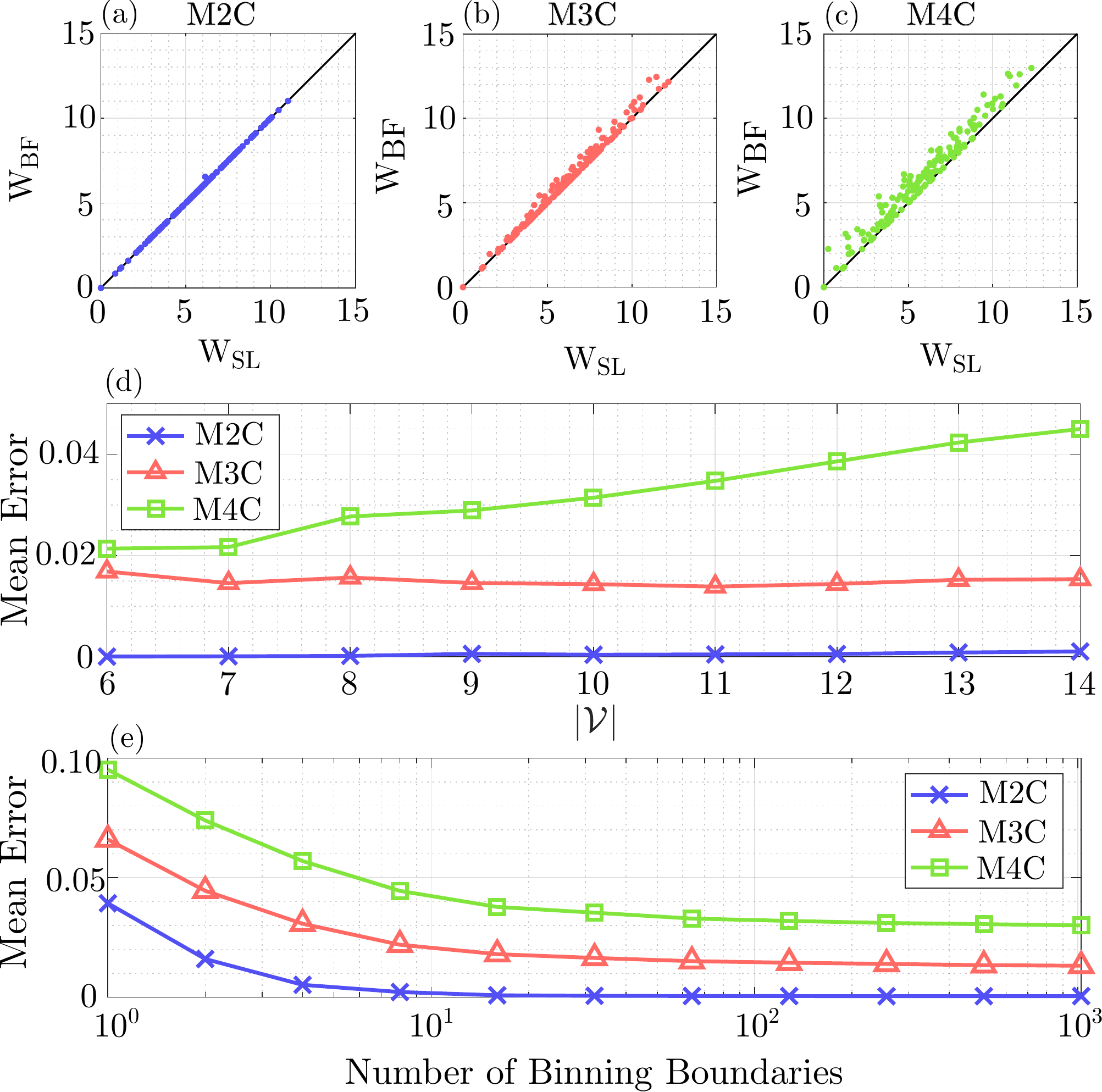}
	\caption{Obtained weights of (a) max-2-cut, (b) max-3-cut and (c) max-4-cut from 160 random graphs using the Stuart-Landau model ($W_{SL}$) with $|\mathcal{V}| = 10$ and 100 unique binning boundaries and the brute force method ($W_{BF}$). Black line indicates $W_{SL}=W_{BF}$. Comparison of the mean error $\bar{S}_W$ obtained by averaging over 160 random graphs of different max-$k$-cut tasks as a function of (d) network size $|\mathcal{V}|$ (tested for 100 unique binning boundaries) and (e) number of binning boundaries with $|\mathcal{V}|=10$.}
	\label{fig:m2c_vs_m3c}
\end{figure}	

We show the weight of cuts solved using the Stuart-Landau model compared to the brute force method for the M2C and M3C in Fig.~\ref{fig:m2c_vs_m3c}(a,b). In the former case, the error between $W_{SL}$ and $W_{BF}$ is negligible. As there are only two possible phase binnings for the M2C problem, every  oscillator has a window of $180^{\circ}$ in which to be correctly binarized. In the latter case, this segment is reduced to $120^{\circ}$, leading to less room for error in the binning process and thus an increase in error between  $W_{SL}$ and $W_{BF}$.  

Furthermore, as we pointed out in Sec.~\ref{sec.2}, mapping the ground state configurations of spin Hamiltonians to the max-$k$-cut problem requires the spins to belong to the corners of the unit $\mathbb{R}^{k-1}$ simplex. This means that the mutual angle between any two spins is either 0 or a constant value (e.g., for the triangle it is $120^\circ$ and for the tetrahedron it is $\approx 70.53^\circ$). For this reason, projecting the XY ground state spins to quaternary decision variables corresponding to the angles $\{0, \pi/2, \pi, 3\pi/2\}$ on the unit circle, with the possible set of values $\cos{( \theta _i - \theta_j)}\in $ $\left\{  1, 0, -1\right\}$, is expected to perform worse in solving the max-4-cut (M4C). Indeed, Fig.~\ref{fig:m2c_vs_m3c}(c) shows that the weights found by the Stuart-Landau network have a greater spread away from the optimal weight. The difference between the $k=2,3,4$ max-$k$-cuts becomes even more apparent when we plot the normalized mean error in Fig.~\ref{fig:m2c_vs_m3c}(d) as a function of graph size. Interestingly, the error for the M2C stays practically negligible indicating that Stuart-Landau systems can compete with photonic Ising machines. The larger error for the M3C is expected due to the higher partition complexity (i.e., 3 binning options) but amazingly stays practically invariant with graph size in contrast to the growing error of the M4C. We finally point out that even though the error of the M4C is increasing with system size it remained below $\bar{S}_W<0.05$ for $|\mathcal{V}| =14$ vertex graphs which possess $4^{13} = 67108864$ different partitions. This opens an exciting perspective in using the condensate network phase dynamics to go beyond optimizing the M3C even if a direct mapping no longer exists. We also plot the mean error of the different max-$k$-cut problems while scanning the number of binning boundaries tested [Fig.~\ref{fig:m2c_vs_m3c}(e)], which shows that the error converges to some minimum asymptotic value as the number of binning boundaries. Notably, the asymptotic error increases with the number of cuts.

%





\end{document}